\documentclass[pre,showpacs,10pt,amsmath,amssymb]{revtex4}
\usepackage{graphicx}

\def\'{'}

\begin{document}

\title{Nonlinear stochastic equations with multiplicative L\'evy noise}

\author
{Tomasz Srokowski}

\affiliation{
 Institute of Nuclear Physics, Polish Academy of Sciences, PL -- 31-342
Krak\'ow,
Poland }

\date{\today}

\begin{abstract}
The Langevin equation with a multiplicative L\'evy white noise is solved. 
The noise amplitude and the drift coefficient have a power-law form. 
A validity of ordinary rules of the calculus for the Stratonovich 
interpretation is discussed. The 
solution has the algebraic asymptotic form and the variance may assume 
a finite value for the case of the Stratonovich interpretation. The problem 
of escaping from a potential well is analysed numerically; predictions of 
different interpretations of the stochastic integral are compared. 
\end{abstract}

\pacs{02.50.Ey,05.40.Ca,05.40.Fb}

\maketitle

\section{Introduction}

The Langevin equation can not always be expressed by means of a deterministic 
drift term, supplemented by a time-dependent stochastic force (an additive noise). 
A physical quantity, which is represented by the random component, may require 
the noise to depend on the stochastic variable itself. In the Langevin description, 
that dependence emerges as a variable noise amplitude (a multiplicative noise). 
The multiplicative noise emerges also in descriptions of complicated systems, 
as a result of the elimination of fast degrees of freedom. The stochastic equation 
is then of the form 
\begin{equation}
\label{la}
\dot x=F(x)+G(x)\eta(t), 
\end{equation}
where $F(x)$ and $G(x)$ are given functions. The stochastic force $\eta(t)$ is 
uncorrelated, $\langle\eta(t)\eta(t')\rangle=\delta(t-t')$, and it is characterised 
by a given probability distribution. In the present paper we assume that $\eta$ 
has the symmetric stable L\'evy distribution defined by the Fourier transform  
\begin{equation}
\label{levd}
\widetilde{L}_\alpha(k)=\exp(-|k/\sigma|^\alpha), 
\end{equation}
where $\alpha$ is the order parameter and $\sigma$ scales the distribution. In Secs.II 
and III we assume $\sigma=1$. 
The case $\alpha=2$ corresponds to the normal distribution which is well known in the 
context of multiplicative processes \cite{vkam,schen}. 

The general and stable L\'evy processes, for $\alpha\ne 2$, exhibit 
long tails of the distribution which makes the moments divergent. They are frequently 
encountered in nature, since long jumps are associated with a complex structure 
of the environment, in particular with long-range correlations. Examples can be 
found in biological physics \cite{wes}, disordered media \cite{bou} and finance 
\cite{car1,man1,san}. 
A master equation description of thermal activation of particles within 
the folded polymers \cite{bro} also involves the multiplicative L\'evy noise 
in a sense that the equation is fractional (L\'evy jumps) 
and it contains a variable diffusion coefficient. 
However, a direct Langevin representation of the topological complexity problem 
is unknown \cite{garb}. 
Since the complex environment is usually nonhomogeneous, one can expect that 
the L\'evy noise in the Langevin equation is rather multiplicative than additive. 
Therefore formalisms, which are supposed to describe complex processes and 
which do that in terms of the additive noise alone, 
may miss essential features of the problem. 
For example, in the field of finance, the standard Black-Scholes equation contains 
the additive Gaussian noise. Eq.(\ref{la}) in its general form 
could be an important generalisation of that equation \cite{car1,man1}. 

Eq.(\ref{la}) is not sufficiently defined for the white noise because it is not clear 
at which time $G(x(t))$ should be evaluated. In the following, we define the stochastic 
integrals, connected with Eq.(\ref{la}), as Riemann integrals. 
According to Stratonovich, one assumes 
\begin{equation}
\label{strat}
\int_0^t G[x(\tau)]d\eta(\tau)=
\sum_{i=1}^n G\left[\frac{x(t_{i-1})+x(t_i)}{2}\right][\eta(t_i)-\eta(t_{i-1})],  
\end{equation}
where $t=\tau n$ and $\tau=t_i-t_{i-1}$ is a time step. 
This interpretation is appropriate for many physical phenomena since it constitutes 
a white noise limit for correlated processes. In this case Eq.(\ref{la}) can be solved 
like usual differential equation, which can be rigorously proved if $\eta$ has the 
convergent variance \cite{gar,zee}. In particular, one can introduce a transformation 
\begin{equation}
\label{tran}
y(x)=\int_{x_0}^x\frac{dx'}{G(x')},~~~~~~~~\hat F(y)=F(x(y))\frac{dy}{dx}, 
\end{equation}
which leads to the Langevin equation with the additive noise: 
\begin{equation}
\label{las}
\dot y=\hat F(y)+\eta(t). 
\end{equation}
Alternatively, we can simply assume 
\begin{equation}
\label{ito}
\int_0^t G[x(\tau)]d\eta(\tau)=\sum_{i=1}^n G[x(t_{i-1})][\eta(t_i)-\eta(t_{i-1})], 
\end{equation}
which formula defines the It\^o interpretation. 
Predictions of Eq.(\ref{la}) in both interpretations are different but in the case 
$\alpha=2$ in Eq.(\ref{levd}) there is a simple relation between them: 
the difference resolves 
itself to the spurious drift \cite{gar}. For $\alpha<2$ such a relation does not exist. 
The Stratonovich interpretation predicts a dependence of the probability distribution 
on the noise amplitude which may change the diffusion properties of the system; 
in particular the accelerated diffusion, in the case of the additive noise, can change 
to the subdiffusion. That problem is discussed in Ref.\cite{sro} for the case without 
drift and for the linear drift. In the It\^o interpretation, in turn, shape of 
the distribution tail is not affected by the amplitude \cite{sro2}. 

In this paper we discuss properties of the Langevin equation which is driven by 
the multiplicative L\'evy noise and nonlinear forces, in particular the problem 
of escaping from a potential well. In Sec.II, 
properties of stochastic integrals for the stable L\'evy processes 
and those with truncated distributions are compared. The Fokker-Planck equation 
for the problem of an algebraic, nonlinear potential is solved in Sec.III. 
The escape from the potential well, understood as the first passage time problem, 
is calculated in Sec.IV and results for both interpretations of the stochastic 
integral are compared.

\section{Stable L\'evy distributions versus truncated ones}

A well known property of the Stratonovich integral (\ref{strat}) allows us to apply 
standard rules of the calculus and then to reduce Eq.(\ref{la}) to an equivalent 
equation with the additive noise. It can be proved \cite{gar,zee} 
for the normally distributed 
noise, i.e. on the assumption that increments are independent and the variance 
is finite. For the general L\'evy stable processes the latter condition is not 
satisfied. However, we can approximate L\'evy distributions by introducing 
a truncation at some large value of the argument either in a form of the sharp 
cut-off or as a rapidly falling tail. Then a sum of stochastic variables 
converges to the normal distribution, according to the central 
limit theorem. Since in the physical phenomena 
process values are usually finite, introducing truncated distributions is realistic. 
In the random walk theory, the truncated L\'evy flights are often considered 
\cite{car1,man,man1,kop,sro1}. 
They agree with the L\'evy flights for an arbitrarily large jump value; 
deviations appear only at very far tails \cite{man}.  
However, there are also remarkable differences between processes which involve 
the stable distribution and the truncated one. We will demonstrate that difference 
for a simple case of the linear noise. 

Let $F(x)=0$ and $G(x)=x$. If the standard rules of the calculus work -- 
we can expect that for the Stratonovich interpretation -- the variable 
in Eq.(\ref{la}) can be changed. As a result we obtain 
from Eq.(\ref{las}) the probability density distribution in the 'log-L\'evy' form 
\begin{equation}
\label{logl}
p(x,t)=\frac{1}{|x|}L_\alpha(\ln(x/x_0),t),
\end{equation}
where $L_\alpha$ denotes the L\'evy distribution with order parameter $\alpha$, 
width parameter $t$ and $x_0=x(0)$. If the process is continuous the point 
$x_0$ acts as an absorbing barrier, i.e. $x>0$ $(x<0)$ for $x_0>0$ $(x_0<0)$. 
It is the case for the Wiener process but it may no longer be true if 
the variance is divergent; then Eq.(\ref{logl}) is no longer valid. 
The distribution in the form Eq.(\ref{logl}) for $\alpha=2$ is known as 
the log-normal distribution and it is 
frequently encountered in nature, e.g. electron velocities in the solar wind \cite{shlc}, 
as well as rainfall amounts \cite{atc} obey this statistics. Moreover, it can serve as 
a natural model of the multifragmentation \cite{shlw}. 

On the other hand, we can solve Eq.(\ref{la}) for the Stratonovich interpretation 
directly from the definition, by means of Eq.(\ref{strat}). 
The discretisation gives us $x_2=x_1+(x_1+x_2)\eta_1\tau^{1/\alpha}/2$; 
therefore $x_2=x_1(1+a_1)/(1-a_1)$, where $a_1=\eta_1\tau^{1/\alpha}/2$. 
The final solution of the stochastic equation reads 
\begin{equation}
\label{slog}
x(t)=x_0\prod_{i=1}^n\frac{1+a_i}{1-a_i}, 
\end{equation}
where $n=t/\tau$. If a cut-off is introduced, $a_i\ll 1$ for a small $\tau$. We take the 
logarithm of Eq.(\ref{slog}), approximate $\ln(1+x)$ by $x$ and neglect terms of 
the order $\tau^2$ and higher. That procedure yields
\begin{equation}
\label{slogtr}
\ln(x(t)/x_0)=\tau^{1/\alpha}(\eta_1+\eta_2+\dots+\eta_n). 
\end{equation}
The above expression converges to the L\'evy distribution with the order parameter 
$\alpha$, unless $x$ is large compared to the cut-off position, and we obtain 
Eq.(\ref{logl}). Since the variance is finite for the process with the truncated 
distribution, in the limit $n\to\infty$ the normal distribution must be reached, 
according to the central limit theorem, but the convergence is extremely slow. 

The case of the stable distribution is distinguished by the presence a 
considerable number of events for which $a_i$ is not small for any given 
$\tau$. Difference in respect to the case of the truncated distribution, 
due to the presence of those events, becomes visible when we consider 
the distribution $p(x,t)$ for negative $x/x_0$. 
Obviously in the limit of small $\tau$, $x(t)/x_0>0$ if any cut-off is introduced. For 
the case without truncation, $x(t)/x_0$ may turn to the negative. Let us estimate the 
probability $P$ that this cannot happen, i.e. that all factors in Eq.(\ref{slog}) 
are positive. We have $P=(p_i)^n$, where 
$p_i=\int_{-\infty}^1 p(a_i)da_i=\int_{-\infty}^{2/\tau^{1/\alpha}}p(\eta)d\eta$. 
Since $\tau\to0$, we may insert the asymptotic form of the L\'evy distribution, 
$p(\eta)=1/|\eta|^{1+\alpha}$ $(|\eta|\gg 1)$. Then 
$p_i=1-\int_{2/\tau^{1/\alpha}}^\infty pd\eta\approx 1-\tau/\alpha 2^\alpha$. 
Finally we have
\begin{equation}
\label{plog}
P=\left(1-\frac{\tau}{\alpha 2^\alpha}\right)^n\longrightarrow \exp\left(-\frac{t}
{\alpha 2^\alpha}\right). 
\end{equation}
Probability that at least one of the terms in Eq.(\ref{slog}) becomes negative, $1-P$, 
appears finite and it rises with time to unity. 
One can easily demonstrate that $P$ converges to one with $\tau\to0$ for any $t$ if 
$p(\eta)$ is normally distributed. In this case $p(x,t)=0$ for $x<0$. 
\begin{figure}[tbp]
\includegraphics[width=8.5cm]{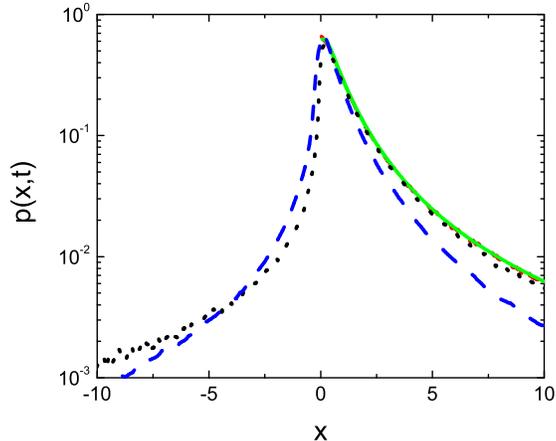}
\caption{(Colour online) Probability distribution, calculated from 
Eq.(\ref{logl}), for the case $G(x)=x$ and $F(x)=0$ with $\alpha=1.5$ at $t=1$, compared 
with the distribution calculated according to Eq.(\ref{slog}) with a truncation at a large 
value of the noise, such that $a_i>0$ (upper curves for $x>0$: solid green line 
and dashed red line, respectively). Those distributions are 
identical. The distribution which follows from Eq.(\ref{slog}), but without any truncation, 
is marked by the dotted black line. The result of the It\^o interpretation 
is also shown (dashed blue line). 
}
\end{figure}

Numerical analysis of the above case is presented in Fig.1. The distribution (\ref{logl}) 
was evaluated by means of the series expansion 
\begin{equation}
\label{sze}
L_\alpha(x,t)=\frac{1}{\pi t^{1/\alpha}\alpha}\sum_{n=0}^\infty
\frac{\Gamma[1+(2n+1)/\alpha]}{(2n+1)!!}(-
1)^n\left(\frac{x}{t^{1/\alpha}}\right)^{2n}.
\end{equation}
The result for the truncated distribution is identical with Eq.(\ref{logl}) 
whereas the case without any cut-off (marked by dots) exhibits a branch 
for the negative $x$. However, both distributions for $x>0$ are very similar and then 
Eq.(\ref{logl}) can serve as an approximation of the L\'evy stable case. 
The result for the It\^o interpretation, Eq.(\ref{ito}), is also presented in Fig.1. 
It falls much faster than the Stratonovich one. 

\section{Nonlinear case}

In this section we consider stochastic processes which are governed by Eq.(\ref{la}) 
with a nonlinear deterministic force. This problem is an important generalisation, 
compared to the linear case, since the corresponding Newton equation may become 
nonintegrable and the dynamics is then chaotic. It happens for a periodic time-
dependent driving (the Duffing oscillator) or if the system has more than 
two degrees of freedom \cite{lich}. We assume the algebraic $F(x)$ and $G(x)$: 
\begin{equation}
\label{fg} 
F(x)=-|x|^\gamma \hbox{sgn}x~~~~~~~~\mbox{and}~~~~~~~~~G(x)=|x|^{-\theta/\alpha}. 
\end{equation}
In the new variable,
\begin{equation}
\label{yodx}
y(x)=\frac{\alpha}{\alpha+\theta}|x|^{(\alpha+\theta)/\alpha}\hbox{sgn}x, 
\end{equation}
the Langevin equation, Eq.(\ref{las}), has the additive noise. 
The corresponding fractional Fokker-Planck equation is of the form 
\begin{equation}
\label{fps0}
\frac{\partial}{\partial t}p(y,t)=K\frac{\partial}{\partial y}|y|^\beta \mbox{sgn}(y)
p(y,t)+\frac{\partial^\alpha}{\partial |y|^\alpha}p(y,t), 
\end{equation}
where $\beta=1-(1-\gamma)(1+\theta/\alpha)$ and $K=(1+\theta/\alpha)^\beta$. 
The drift term in Eq.(\ref{fps0}) corresponds to the effective potential 
$V(y)\sim|y|^{(\gamma-1)(1+\theta/\alpha)}$. 
We are interested in the asymptotic shape of a steady-state solution $p_S(x)$. 
The solution for large $|y|$ can be found by taking into account small wave numbers 
in the Fourier expansion. The Fourier transform of Eq.(\ref{fps0}) in the stationary 
limit reads
\begin{equation}
\label{kfps0}
Kk\frac{\partial}{\partial k}{\cal F}(|y|^{\beta-1}p_S(y))=|k|^\alpha{\widetilde p}_S(k). 
\end{equation}
We assume the solution in the form of the Fox function \cite{mat,sri}, 
\begin{eqnarray} 
\label{sols}
p_S(y)=NH_{2,2}^{1,1}\left[|y|
\left|\begin{array}{l}
(a_1,A_1),(1/2,1/2)\\
\\
(0,1),(b_2,1/2)
\end{array}\right.\right],  
\end{eqnarray}
where $N$ is the normalisation constant and the coefficients are to be determined. 
Some useful properties of the Fox functions are presented in Appendix. 
Eq.(\ref{sols}) represents the stable and symmetric 
L\'evy distribution for $a_1=1-1/\alpha$, $A_1=1/\alpha$ and $b_2=1/2$ \cite{sch}. 
We insert Eq.(\ref{sols}) into Eq.(\ref{kfps0}) and apply the general formula (\ref{A.1}) 
in order to get rid of the algebraic factor. Then we calculate the Fourier transform, 
according to the formula (\ref{A.2}), 
and expand both sides of Eq.(\ref{kfps0}) by using the Fox function series 
representation, Eq.(\ref{A.3}). Eq.(\ref{kfps0}) takes the form
\begin{equation}
\label{szer} 
Kk\frac{\partial}{\partial k}\left(c_1+c_2|k|^{w_1}+c_3|k|^{w_2}+o(k^2)\right)=
|k|^\alpha[1+o(|k|^\alpha)],
\end{equation}
where $w_1=(1-a_1)/A_1-\beta$, $w_2=(2-a_1)/A_1-\beta$ and $c_i$ are constants. The above 
equation is satisfied if $w_1=0$ and $w_2=\alpha$ which conditions determine the coefficients: 
$a_1=1-\beta/\alpha$ and $A_1=1/\alpha$. The condition $K\alpha c_3=1$, where 
\begin{equation}
\label{c3}
c_3=N\frac{\alpha(\alpha+1)}{2\pi}\frac{\Gamma(-\alpha)\Gamma(\alpha+\theta+1)\cos(\pi\alpha/2)}
{\Gamma(1+(\alpha-\theta)/2)\Gamma(-b_2+(1-\alpha+\theta)/2)}, 
\end{equation}
can be satisfied by an appropriate choice of $b_2$. 

The asymptotic behaviour of $p_s(y)$ follows from expansion of the Fox function 
in powers of $|y|^{-1}$: it can be obtained by a variable transformation 
$y\to y^{-1}$ by means of Eq.(\ref{A.4}) and by applying Eq.(\ref{A.3}). The first term 
produces the result $p_s(y)\sim |y|^{-\alpha-\beta}$ 
$(|y|\to \infty)$ which, after transformation to the original variable according to 
the formula $p_s(x)=p_s(y(x))|dy/dx|$, yields the final result
\begin{equation}
\label{asym}
p_s(x)\sim |x|^{-(\alpha+\theta+\gamma)}~~~~~~(|x|\to \infty). 
\end{equation}
To satisfy the normalisation condition, we assume $\alpha+\theta+\gamma>1$. 
Eq.(\ref{asym}) predicts the L\'evy stable distribution 
with a divergent variance for $\alpha+\theta+\gamma<3$. 
If $\alpha+\theta+\gamma\ge3$, the variance is finite though higher moments 
may be divergent. Therefore, long tails of the distribution 
can be confined either by choosing 
a sufficiently steep potential or an appropriate noise. The latter must be 
such that amplitude declines with position sufficiently fast (large $\theta$) 
and/or the order parameter $\alpha$ is large (steep tails). 
The case $\gamma=1$ corresponds to the harmonic 
oscillator; it is discussed in Ref.\cite{sro}. If $\theta=0$ and $\gamma$ is an odd integer, 
an analytical expression for the stationary probability distribution, valid for arbitrary $|x|$, 
can be derived \cite{che}. 
\begin{figure}[tbp]
\includegraphics[width=8.5cm]{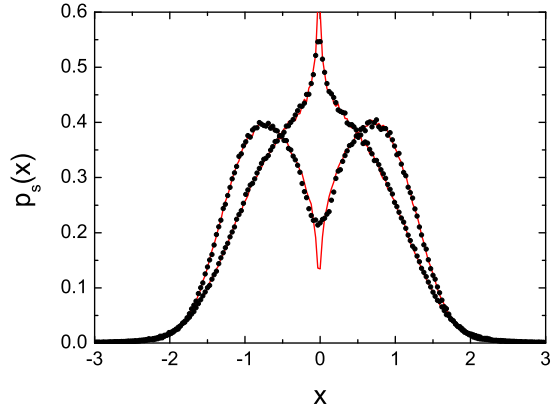}
\caption{(Colour online) Stationary probability distributions, for the system 
given by Eq.(\ref{fg}), calculated by applying the 
transformation (\ref{yodx}) (lines) and by using Eq.(\ref{disc}) (points) for 
$\alpha=1.8$, $\gamma=2.5$ and two values of $\theta$: -0.2 (the case with a maximum 
in the origin) and 0.5.}
\end{figure}

On the other hand, we solve Eq.(\ref{la}) by a numerical simulation 
of stochastic trajectories. 
It can be performed in two ways. First, we directly apply the discretisation 
formula which follows from Eq.(\ref{strat}) and is of the form \cite{wer} 
\begin{equation}
\label{disc}
x_{i+1}=x_i+F(x_i)\tau+[G(x_i)+G(x_{i+1})]\eta_i\tau^{1/\alpha}/2. 
\end{equation}
To find the process value, one has to solve, at each step, the following 
nonlinear equation
\begin{equation}
\label{rnax2}
x_{i+1}-a_iG(x_{i+1})-x_i-F(x_i)\tau-G(x_i)a_i=0, 
\end{equation}
where $a_i=\eta_i\tau^{1/\alpha}/2$. For that purpose we apply the parabolic interpolation 
scheme (the Muller method) \cite{ral}. The algorithm must be carefully implemented since, 
due to the explicit multiplication of the noise by the $x-$dependent 
factor, the round-off errors are large and then it is difficult 
to achieve a high accuracy of the results. Alternatively, we can first transform 
Eq.(\ref{la}) to Eq.(\ref{las}), simulate trajectories $y(t)$ to find $p(y)$ 
and finally transform the distribution: $p(y)\to p(x)$. 
Comparison of both algorithms is presented in Fig.2 for positive and negative $\theta$; 
the distributions are actually identical. In the following simulations we apply 
the method of variable transformation. 
\begin{figure}[tbp]
\includegraphics[width=8.5cm]{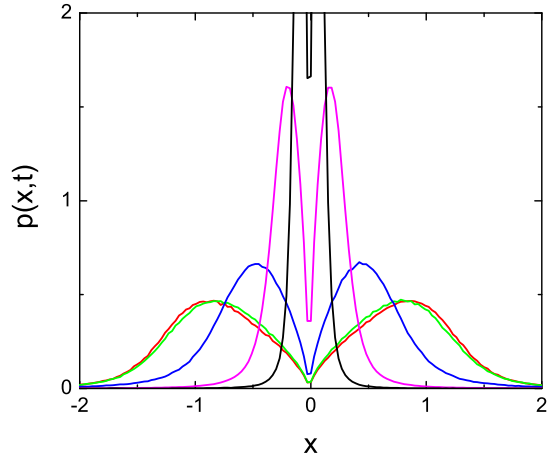}
\caption{(Colour online) Time evolution of the probability distribution in the case  
Eq.(\ref{fg}) for 
$\alpha=1.5$, $\theta=1$ and $\gamma=2.5$. The distribution was evaluated at the 
following times: 0.001, 0.01, 0.1, 0.5, and 1, which cases correspond to the rising width. 
The steady state is reached at $t=1$. 
}
\end{figure}

Distributions which are initially positioned at $x=0$ evolve with time to the steady state. 
The convergence appears very fast. Example of the evolution is presented in Fig.3 
for $\alpha=1.5$ and $\theta=1$. The stationary distribution is reached already at $t=1$. 
\begin{figure}[tbp]
\includegraphics[width=8.5cm]{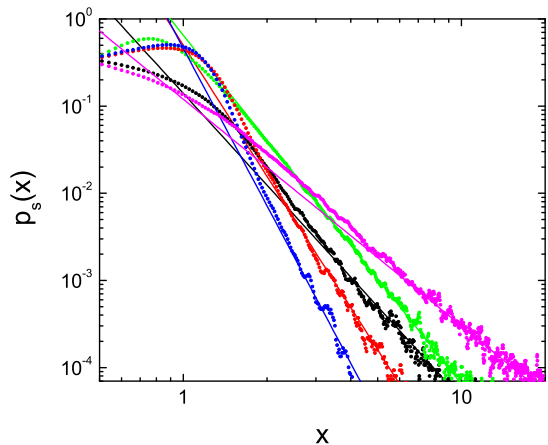}
\caption{(Colour online) Stationary distributions, calculated numerically, for the system 
given by Eq.(\ref{fg}) (points). The following cases are presented: 1. $\gamma=1.6$, 
$\alpha=1.5$, and $\theta=-0.5$; 2. $\gamma=2.5$, $\alpha=0.5$, and $\theta=1$; 3. $\gamma=2.5$, 
$\alpha=1.5$, and $\theta=-0.5$; 4. $\gamma=2.5$, $\alpha=1.5$, and $\theta=1$; 
5. $\gamma=3.5$, $\alpha=1.5$, and $\theta=1$; (from top to bottom 
at the right hand side). Slopes of the straight lines follow from Eq.(\ref{asym}).  
}
\end{figure}

Various sets of parameters $\alpha$, $\theta$ and $\gamma$ define processes which are either 
stable L\'evy ones, with divergent variance, or 
processes with heavy tails, for which the variance 
exists but higher moments are divergent. Examples are presented
in Fig.4. The case of negative $\theta$ and a weakly changing $F(x)$ 
(the case 1. in the figure) corresponds to the 
slope 2.6, in the other cases the slope is larger than three. Slopes of the straight lines 
in the figure follow from the asymptotic formula, Eq.(\ref{asym}), and they agree with 
the numerical results. 

\section{Escape from a potential well}

A particular case of Eq.(\ref{la}), which involves the nonlinear deterministic force and 
the boundary conditions, is the problem of passing over a potential barrier. 
This problem is of great physical importance and it has been 
extensively studied for the case of the normal distribution \cite{han}. For example, fusion 
of heavy ions in nuclear physics consists in a transfer of mass over the 
Coulomb barrier. A multiplicative noise emerges when one considers a parametric 
activation of the potential, i.e. if height of the barrier randomly varies 
\cite{woz}. Increasing intensity of the multiplicative noise in the bistable 
stochastic system can produce a stochastic resonance \cite{gamm}. 
Properties of systems driven by general L\'evy stable noises may be different 
than those for $\alpha=2$. In particular, a waiting time for noise-induced jumping 
between metastable states may depend, due to the presence of single  
long jumps, more on the width than on the height of the barrier \cite{dit}. 
Moreover, the ratio of first mean passage time from one to the other minimum 
is no longer twice of the time to reach the top of the barrier \cite{dyb2}. 
Asymmetry in the L\'evy distribution affects the escape time; 
it can both enhance and suppress the escape events. 
The rate of escape, as a function of the parameter $\alpha$, is discontinuous. 
It was recently demonstrated that a double stochastic resonance can be 
observed in a single well potential without explicit external driving, 
if the L\'evy stable noise is introduced \cite{dyb3}. 
\begin{figure}[tbp]
\includegraphics[width=8.5cm]{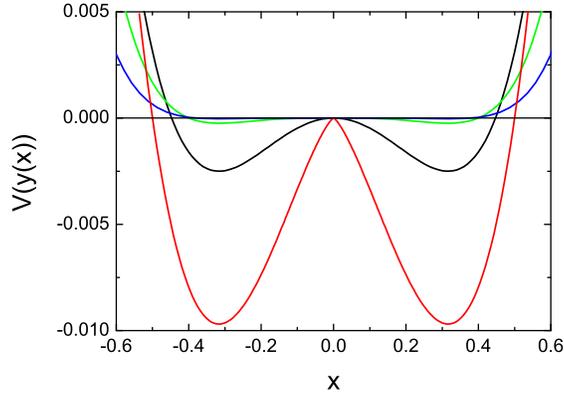}
\caption{(Colour online) The effective potential in the Stratonovich interpretation, 
calculated from Eq.(\ref{drb}), for $A=1$, $B=0.1$ and the following values of 
$\theta/\alpha$: 4/3, 2/3, 0, -1/3 (from top to bottom). 
}
\end{figure}

In this section we 
consider the problem of escaping from the potential well for the multiplicative 
noise and, in particular, the dependence of the mean first passage time (MFPT) on 
the specific interpretation of the stochastic integral, either Stratonovich or It\^o. 
We assume the potential in the form
\begin{equation}
\label{potb}
V(x)=\frac{A}{4}x^4-\frac{B}{2}x^2 
\end{equation}
and the noise amplitude $G(x)$ is given by Eq.(\ref{fg}). The main quantity of 
interest is the dependence of MFPT on the parameter $\theta$, which quantifies 
the noise amplitude variability. 
The transformed Fokker-Planck equation with additive noise is the following
\begin{equation}
\label{fpsb}
\frac{\partial}{\partial t}p(y,t)=-\frac{\partial}{\partial y}F(y)
p(y,t)+\frac{\partial^\alpha}{\partial |y|^\alpha}p(y,t), 
\end{equation}
where
\begin{equation}
\label{drb}
F(y)=(1+\theta/\alpha)y\left[B-A[(1+\theta/\alpha)|y|]^{2\alpha/(\alpha+\theta)}\right].
\end{equation}
The effective drift $F(y)$ depends only on the ratio $\theta/\alpha$. We can infer some 
qualitative conclusions about the dynamics from the shape of an effective potential, 
which follows from Eq.(\ref{drb}). This potential, as a function of the original 
variable $x$, is presented in Fig.5. The height 
of the barrier falls sharply with $\theta/\alpha$ -- the potential is very shallow for 
negative $\theta$ -- whereas position of the barrier is constant.  Therefore 
we can expect a suppression of transport, which is defined by the boundary conditions 
in the variable $x$, for large $\theta/\alpha$. 
\begin{figure}[tbp]
\includegraphics[width=8.5cm]{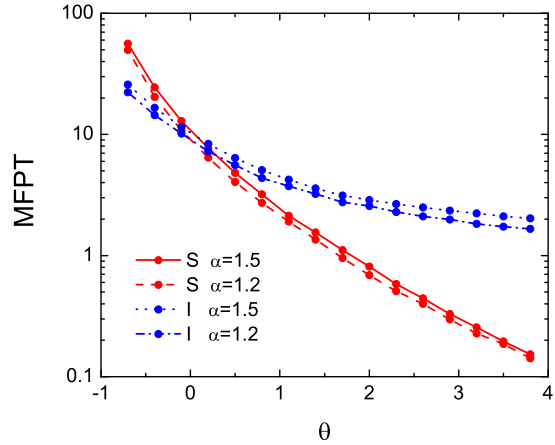}
\caption{(Colour online) Mean first passage time as a function of 
$\theta$ for $\alpha=1.5$ and 1.2, calculated for the potential (\ref{potb}) 
with $A=1$ and $B=0.1$. Results for both interpretations of the stochastic 
integral are presented. 
}
\end{figure}

A numerical analysis of the potential barrier problem must take into account that 
the system under consideration is limited in space. Long tails of 
the distribution may not manifest 
themselves if the available space is too small. Therefore, in the following, 
we rescale the system by putting in Eq.(\ref{levd}) $\sigma=0.1$. 
MFPT, defined as a time the particle needs to pass for the first time from the 
left minimum of the potential to any $x>0$, was calculated by numerical solving 
of Eq.(\ref{las}) with the initial condition $x(0)=-\sqrt{B/A}$ and 
the boundary condition at the absorbing barrier, $x=0$. 
The average time, as a function of $\theta$ for two values of $\alpha$, 1.5 and 1.2, 
is presented in Fig.6. It falls sharply with $\theta$, like an exponential, 
which is a consequence of the shallow effective potential for large $\theta$. 
Results for both values of $\alpha$ are similar, they differ only by a constant factor. 
MFPT rises with $\alpha$ since jumps become shorter. 
This result is presented in Fig.7; a difference between the case of positive 
$\theta$ and negative one is substantial. 
\begin{figure}[tbp]
\includegraphics[width=8.5cm]{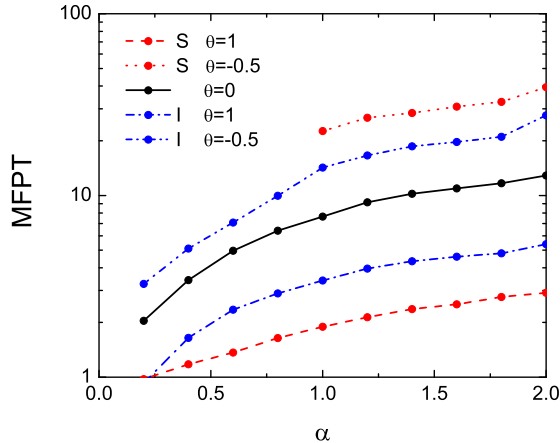}
\caption{(Colour online) Mean first passage time in both interpretations 
of the stochastic integral, as a function of $\alpha$, for some values of $\theta$.
}
\end{figure}

Predictions for the It\^o interpretation are 
also presented in Figs.6 and 7. The dependence of MFPT on $\theta$ is much 
weaker than for the Stratonovich case; MFPT falls algebraically for large 
$\theta$. In this case, the dynamics is affected 
by the noise only near the top of the barrier -- the noise is then 
localised and very strong -- while inside the well the deterministic 
trapping dominates. Results for different $\alpha$ are qualitatively the same. 
Moreover, MFPT rises with $\alpha$ in the It\^o case, 
similarly as in the Stratonovich interpretation.

\section{Summary and conclusions}

L\'evy distribution is characterised by long tails which cause the 
divergent moments. If noise in the Langevin equation 
is defined in terms of the L\'evy distribution, the influence of the tails 
can be confined either by the potential or by the variable noise amplitude. 
Then the process which is described by the Langevin equation may have 
a finite variance. We studied 
such processes by solving the Langevin equation with the 
multiplicative L\'evy noise and nonlinear drifts. The asymptotic shape of 
the stationary probability distribution depends 
both on the noise amplitude, assumed 
in the power-law form with the parameter $\theta$, and on the potential 
slope $\gamma$, according to Eq.(\ref{asym}). If the above parameters are 
large enough, the variance is finite for any order parameter $\alpha$. 
The asymptotic formula (\ref{asym}) is valid only for 
the Stratonovich interpretation; in 
the It\^o interpretation, the distribution is less sensitive on the slope 
of the noise amplitude (the parameter $\theta$). 
For the case without drift and 
with the linear drift, the asymptotic formula is the same as for the additive noise, 
i.e. the dependence on $\theta$ does not appear \cite{sro,sro2}. 
Then the variance is always infinite. 

The difference between both interpretations of the stochastic integral is 
also visible in the problem of escape from the potential well. 
This problem was studied numerically: MFPT was calculated, as a function 
of $\alpha$ and $\theta$. The effective potential, 
which includes the variable diffusion coefficient in the Stratonovich 
interpretation, possesses a high barrier when $\theta$ is negative. As a 
consequence, MFPT rapidly falls with $\theta$. This effect is not observed 
in the It\^o case: MFPT falls with $\theta$ according to a power-law. 
Moreover, MFPT rises with $\alpha$ in both interpretations. The above conclusions 
are valid only if the relative size of the system is large enough to allow 
the long tails of the L\'evy distribution to manifest themselves. It was ensured 
in the calculations by taking a small value of the noise parameter $\sigma$. 

Both analytical and numerical calculations for the case of the Stratonovich 
interpretation were performed by using the statement that rules of the 
ordinary calculus apply and change of variables is possible. 
That statement is exact if the variance is finite, in particular for 
the truncated distribution. Otherwise, the Langevin equation in 
the transformed variables, Eq.(\ref{las}), offers only an approximation 
to Eq.(\ref{la}) since for the case of L\'evy stable processes  
those equations are not strictly equivalent. The approximation is quite 
accurate but one can also encounter qualitative differences. 
We demonstrated, by considering the case of the linear noise, 
that the stochastic variable may change its sign, which is 
forbidden for the case of the normal distribution or if the cut-off is present. 
A possibility to use Eq.(\ref{las}) is of great practical importance. 
It enables us not only to perform analytical calculations 
but also offers a simple numerical tool of much higher 
precision than the direct integration of Eq.(\ref{la}). 

\section*{APPENDIX}

\setcounter{equation}{0}
\renewcommand{\theequation}{A\arabic{equation}} 

In the Appendix, we present properties of the Fox functions which are used in Sec.III. 
The multiplication rule 
  \begin{eqnarray} 
\label{A.1}
x^\sigma H_{pq}^{mn}\left[x\left|\begin{array}{c}
(a_p,A_p)\\
\\
(b_q,B_q)
\end{array}\right.\right]= 
H_{pq}^{mn}\left[x\left|\begin{array}{c}
(a_p+\sigma A_p,A_p)\\
\\
(b_q+\sigma B_q,B_q)
\end{array}\right.\right], 
  \end{eqnarray} 
 where $x>0$, allows us to evaluate products involving algebraic terms. The cosine 
 Fourier transform is given by the following expression 
   \begin{eqnarray} 
\label{A.2}
\hskip-1cm
\int_0^\infty H_{pq}^{mn}\left[x\left|\begin{array}{c}
(a_p,A_p)\\
\\
(b_q,B_q)
\end{array}\right.\right]\cos(kx)dx= 
\frac{\pi}{k}H_{q+1,p+2}^{n+1,m}\left[k\left|\begin{array}{l}
(1-b_q,B_q),(1,1/2)\\
\\
(1,1),(1-a_p,A_p),(1,1/2)
\end{array}\right.\right]. 
  \end{eqnarray}
 Numerical values of the Fox function can be obtained by means of the following 
 series expansion
   \begin{eqnarray}  
\label{A.3}
\hskip-1cm 
H_{pq}^{mn}\left[x\left|\begin{array}{c}
(a_p,A_p)\\
\\
(b_q,B_q)
\end{array}\right.\right]=\sum_{h=1}^m\sum_{\nu=0}^\infty\frac{\prod_{j=1,j\ne
h}^m\Gamma(b_j-B_j\frac{b_h+\nu}{B_h})\prod_{j=1}^n\Gamma(1-a_j+A_j\frac{b_h+\nu
}{B_h})}{\prod_{j=m+1}^q\Gamma(1-b_j+B_j\frac{b_h+\nu}{B_h})\prod_{j=n+1}^p
\Gamma(a_j-A_j\frac{b_h+\nu}{B_h})}\frac{(-1)^\nu x^{(b_h+\nu)/B_h}}{\nu!B_h}, 
  \end{eqnarray}
where $n\ne 0$. The asymptotic expansion results from the property:
  \begin{eqnarray} 
\label{A.4}
H_{pq}^{mn}\left[x\left|\begin{array}{c}
(a_p,A_p)\\
\\
(b_q,B_q)
\end{array}\right.\right]= 
H_{qp}^{nm}\left[\frac{1}{x}\left|\begin{array}{c}
(1-b_q,B_q)\\
\\
(1-a_p,A_p)
\end{array}\right.\right].
  \end{eqnarray}

\end{document}